\pdfoutput=1
\documentclass[cameraready]{Interspeech}
\title{GigaChat Audio: Time-aware Large Audio Language Model}

\author[affiliation={1}]{Aleksandr}{Kutsakov}
\author[affiliation={1}]{Mariia}{Sadovina}
\author[affiliation={1}]{Georgii}{Gospodinov}
\author[affiliation={1}]{Alexandr}{Maximenko}
\author[affiliation={1}]{\\Oleg}{Kutuzov}
\author[affiliation={1}]{Pavel}{Bogomolov}
\author[affiliation={1}]{Fyodor}{Minkin}
\address{
    $^1$ SaluteDevices, Russia
}

\email{\{askutsakov, sadovinama, georgygospodinov, ae.maximenko, olegkutuzov01, bobrosoft98, minkin.fyodor\}@gmail.com}

\keywords{large audio language models, audio question answering, long-form speech processing}

\begin{document}

\maketitle

% the abstract here must exactly match the abstract entered into the paper submission system
\begin{abstract}
    % 1000 characters. ASCII characters only. No citations.
    Temporal grounding in long recordings remains challenging for audio-conditioned LLMs. We present a time-aware audio LLM that answers questions with explicit timestamps over up to 120 minutes of input. Our approach interleaves periodic time markers with continuous audio tokens using large-scale synthetic supervision from a cascaded pipeline. Our model achieves strong temporal-grounding accuracy on short and long benchmarks and supports time-anchored fragment descriptions and summaries. Extensive ablations examine how time representation, marker frequency, tokenization, and duration-mixture design affect accuracy and computational cost. We release model weights and datasets to support further research on time-aware audio understanding, available at https://huggingface.co/ai-sage/GigaChat3.1-Audio-10B-A1.8B.
\end{abstract}

\section{Introduction}

Long recordings such as meetings, podcasts, lectures, and call-center logs are increasingly consumed through interactive interfaces: users ask questions, request summaries, and navigate to specific evidence. In these settings, \emph{temporal grounding} is not a cosmetic feature but a verifiability primitive: a system should answer not only \emph{what} happened but also \emph{when} it happened, enabling users to jump to the supporting audio segment.

Recent audio-conditioned LLMs enable instruction-following directly from speech and general audio, including both proprietary \cite{gpt4o, gemini2_5} and open models \cite{voxtral, qwen3omni}. However, temporal grounding in long-form audio remains unreliable: models often produce plausible content while emitting non-parseable timestamps, overly coarse time references, or unsupported temporal claims. The core challenge is that time is not naturally represented in standard audio token streams, and long recordings exacerbate the mismatch between the model's internal representation and user-facing timestamped outputs.

This work asks three practical questions: (i) What is the minimal and robust way to represent time in an audio LLM input/output interface? (ii) How should temporal supervision be generated at scale for long recordings, where manual annotation is prohibitively expensive? (iii) Do temporal models trained on a single duration regime generalize to other lengths? Our experiments show strong asymmetry: training only on short audio does not extrapolate to long recordings, while training only on long audio degrades short-audio performance (Figure~\ref{fig:extrap}). Moreover, periodic temporal anchors are essential for long-form grounding: removing them collapses long-recording accuracy (Table~\ref{tab:main}), although in practice even sparse anchors (e.g., once per minute) are sufficient (Table~\ref{tab:freq}).

We present GigaChat Audio, a \emph{time-aware} audio-conditioned LLM that supports up to 120 minutes of input and produces time-anchored answers, fragment descriptions, and summaries with explicit timestamps. Our approach interleaves continuous audio tokens with periodic \emph{inter-timings} that act as temporal anchors. To train temporal behavior at scale, we introduce a cascaded synthetic-data pipeline that generates supervision from timestamped transcripts using slicing to reduce front-loading bias and a global verifier to enforce consistency.

\textbf{Contributions.}
\begin{itemize}
    \item We release an open-weight time-aware Audio LLM for up to 120-minute inputs.
    \item We release a new \textbf{10k+ hours} temporal dataset spanning seconds-to-hours recordings with supervision for temporal QA and time-anchored summarization.
    \item We provide a high-quality long-form synthetic data pipeline with transcript slicing, verification, and robust evaluation by multi-sampling aggregation.
    \item Through extensive ablations, we show that (a) temporal models must be trained on a mixture of audio lengths to generalize, and (b) temporal anchors are necessary, with anchor frequency/format controlling the accuracy--compute trade-off.
\end{itemize}

\begin{figure}[t]
  \centering
  \includegraphics[width=\linewidth]{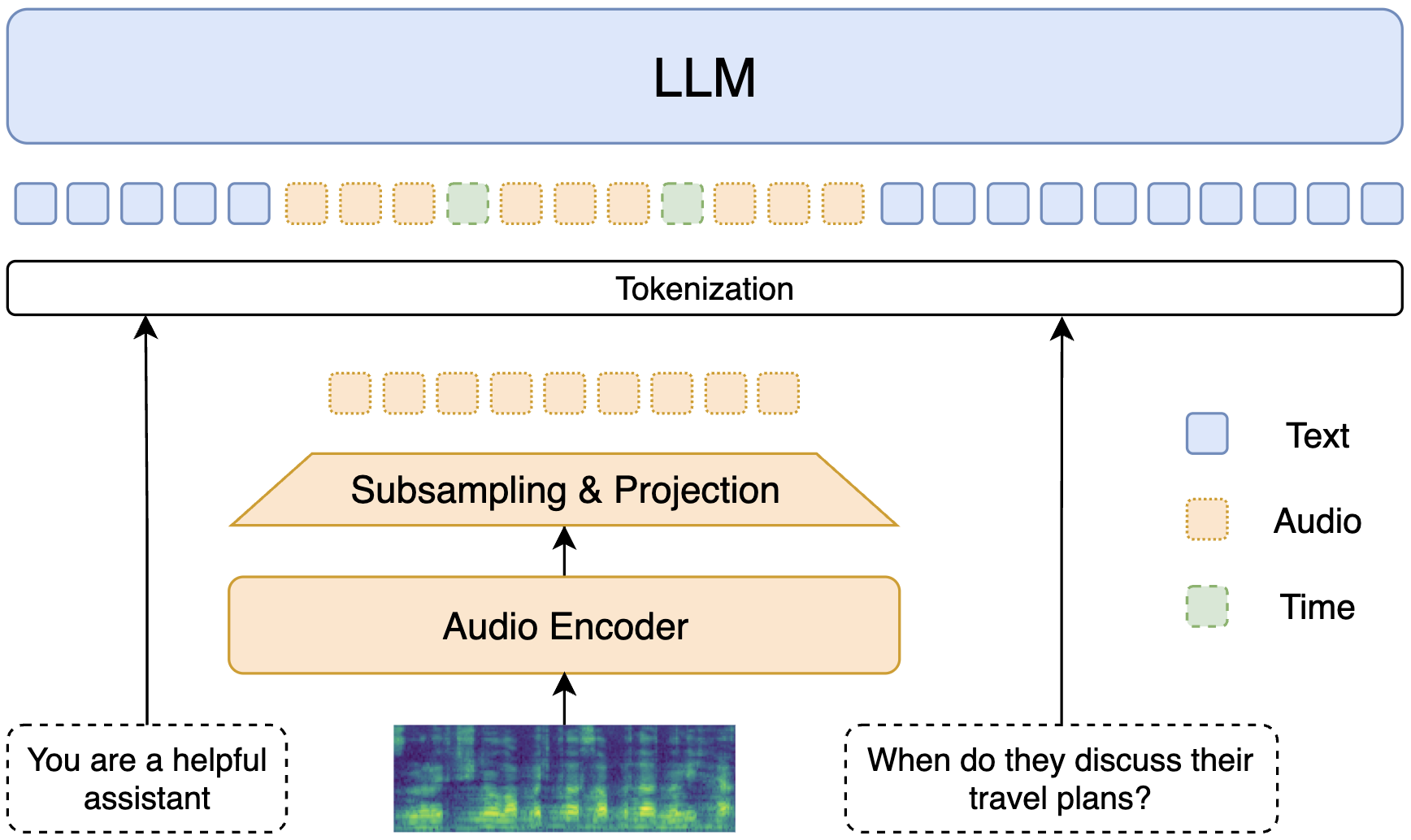}
  \caption{
    Architecture of a time-aware audio LLM.
    Token streams are interleaved: text tokens, continuous audio tokens, timing tokens (either text-form \texttt{hh:mm:ss} or special ones).
  }
  \label{fig:arch}
\end{figure}

\begin{figure*}[!t]
  \centering
  \includegraphics[width=\textwidth]{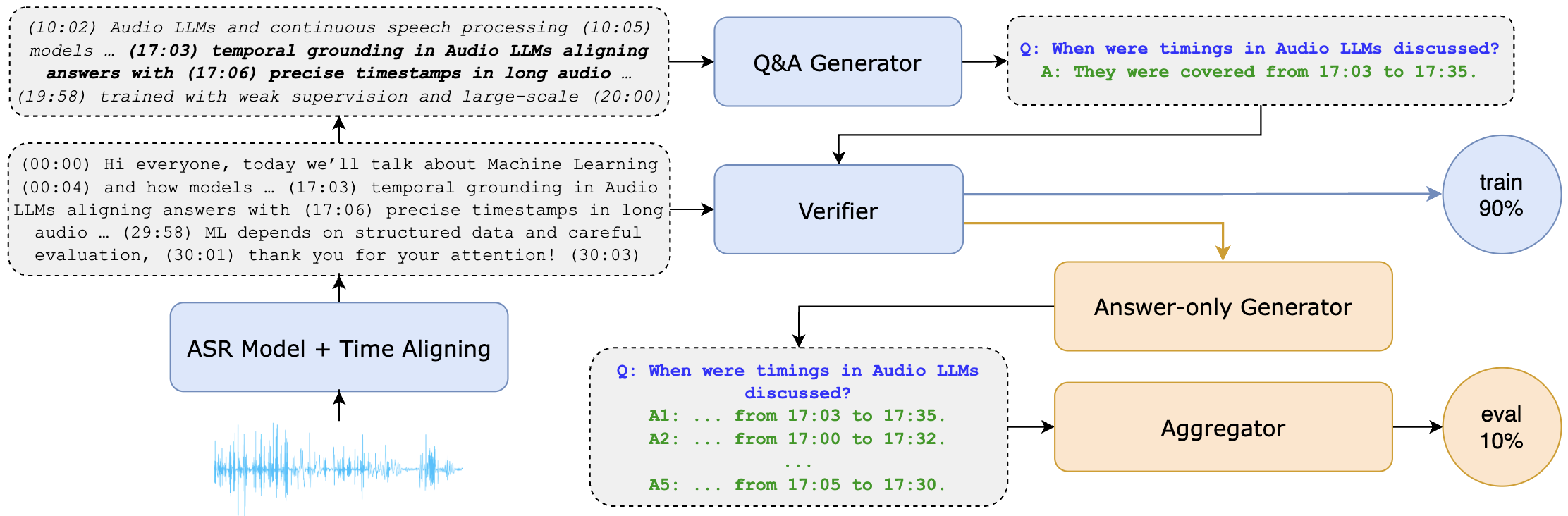}
  \caption{
    Synthetic supervision pipeline for temporal grounding.
    Audio is transcribed and aligned to obtain a timestamped transcript. For QA-style data, we sample a $\sim$10-minute transcript fragment and generate (Q, A) pairs; a verifier checks each pair against the full timestamped transcript to enforce global consistency.
    We split train/eval at the \emph{audio level} (all questions from the same recording are in the same split). For eval, we sample 5 answer variants for each question with an answer-only generator, then aggregate and filter.
  }
  \label{fig:pipeline}
\end{figure*}

\section{Related Work}

Modern multimodal systems can follow instructions from speech and audio inputs, including GPT-4o \cite{gpt4o}, Gemini \cite{gemini2_5}, and open-weight omni models such as Qwen3-Omni \cite{qwen3omni} and Voxtral \cite{voxtral}. Earlier open AudioLLMs \cite{qwen2_audio, salmonn} demonstrate broad audio understanding and dialogue capabilities. While these models support transcription and general audio QA, their temporal grounding behavior - especially on long-form recordings - is rarely isolated and systematically analyzed.

Several systems target long-form ASR and structured rich transcription with explicit speaker and timing information, e.g., VibeVoice-ASR \cite{vibe_voice_asr} and MOSS Transcribe Diarize \cite{moss_transcribe}. Tools such as WhisperX provide word-level timestamp alignment for long recordings \cite{whisperx}. These works focus primarily on transcription-centric objectives, whereas our goal is open-ended question answering and summarization that are explicitly anchored in time.

Temporal grounding has been studied for short audio events via text-to-audio grounding benchmarks \cite{audiogrounding, dcase_qa}. Recent LALM-oriented approaches, e.g., TimeAudio \cite{time_audio}, introduce explicit temporal modeling for audio localization. Related retrieval-style formulations include audio moment retrieval (AMR) and datasets such as Clotho-Moment and CASTELLA \cite{amr, castella}. In contrast, we focus on long recordings (up to 120 minutes) and study practical design choices for representing time (anchor frequency, format, and tokenization) as well as duration-mixture training for length generalization.

Natural-language moment localization in video provides useful inspiration for temporal reasoning and localization \cite{video_moment_localization}, and recent Video-LLMs incorporate temporal separator tokens to improve localization \cite{timemarker, longcat}. For evaluating open-ended generation, LLM-as-a-judge protocols have been widely explored \cite{geval}. We adopt a judge-based evaluation for timed summaries and fragment descriptions, while ensuring temporal grounding is verifiable via interval-overlap metrics.

\section{Method}

\subsection{Model}

Our model is a \emph{time-aware} audio LLM built by attaching an audio front-end to a 10B-A1.8B MoE text checkpoint with a 256k-token context \cite{10b} and fine-tuning it on audio SFT data with explicit time signals (Figure~\ref{fig:arch}). Training is performed using data parallelism only, without tensor or sequence parallelism.

We use an \emph{encoder $\rightarrow$ subsampler $\rightarrow$ projector} audio stack with FlashAttention \cite{flash} and chunk-wise attention in the encoder (8\,s chunks with 40\,ms stride) that produces continuous audio embeddings aligned to the text embedding space at a 160\,ms frame rate. For audio encoder pretraining, we follow the HuBERT-like \cite{hubert} setup of our previous work \cite{gigaam} with the same hyperparameters, using KMeans distributions from a previously trained audio-LLM encoder as distillation targets over $2$M hours of unlabeled multilingual audio. During audio SFT we fix $lr_{\text{dec}} = 5 \cdot 10^{-6}$ and $lr_{\text{enc}} = 10^{-4}$. To make time explicit, we interleave \emph{inter-timings} between continuous ``audio tokens'' either as plain-text timestamps (\texttt{hh:mm:ss}) or as dedicated timing tokens (see Section \ref{subsec:extratokens}), inserted every 60 seconds (see Section \ref{subsec:interfreq} on timing frequency). We also append a final inter-timing marker at the end of each audio sequence.

\begin{table*}[!t]
  \caption{
    \textbf{Main comparison across tasks and datasets.}
    We evaluate AudioGrounding (AGr), AMI Corpus meeting understanding (AMI), time-aware DCASE Audio QA (DAQA), temporal grounding and timed descriptions across duration buckets, and long-form timed summarization.
    We compare against open-source (Qwen3-Omni-30B-A3B, TimeAudio) and proprietary (Gemini 3 Flash) models, and include ablations without inter-timings and on their frequency to isolate the effect of periodic temporal anchors.
  }
  \centering
  \small
  \setlength{\tabcolsep}{4pt}
  \begin{tabular}{lcccccccccccc}
    \toprule
    \textbf{Model} &
    \textbf{AGr ($\uparrow$)} &
    \textbf{AMI ($\downarrow$)} &
    \textbf{DAQA ($\downarrow$)} &
    \multicolumn{3}{c}{\textbf{TGr (mIoU $\uparrow$)}} &
    \multicolumn{3}{c}{\textbf{Descriptions ($\uparrow$)}} &
    \multicolumn{3}{c}{\textbf{Summ (20--40m)}} \\
    \cmidrule(lr){5-7}
    \cmidrule(lr){8-10}
    \cmidrule(lr){11-13}
    &
    \textbf{7--10s}
    &
    \textbf{15--50m}
    &
    \textbf{6--10s}
    &
    \textbf{0--1m} & \textbf{2--5m} & \textbf{20--40m} &
    \textbf{0--1m} & \textbf{2--5m} & \textbf{20--40m} &
    \textbf{Tm} ($\uparrow$) & \textbf{AES} ($\uparrow$) & \textbf{Rd} ($\downarrow$) \\
    \midrule
    Qwen3-Omni-30B &
    50.8 &
    290.5 &
    1.00 &
    47.2 & 27.3 & 3.6 &
    \textbf{3.95} & 3.70 & 2.11 &
    43.7 & 74.3 & 24.9 \\
    TimeAudio &
    \textbf{58.8} &
    -- &
    \textbf{0.12} &
    19.6 & -- & -- &
    1.41 & -- & -- &
    -- & -- & -- \\
    Gemini 3 Flash &
    41.7 &
    \textbf{1.00} &
    0.9 &
    39.3 & 30.2 & 56.1 &
    3.63 & 3.05 & 3.75 &
    73.6 & \textbf{91.3} & \textbf{8.2} \\
    \textbf{Ours (inter=60s)} &
    45.1 &
    3.50 &
    1.70 &
    40.6 & 53.0 & 53.8 &
    3.63 & 3.76 & 3.83 &
    76.7 & 88.1 & 16.9 \\
    \quad w/ inter=7s &
    39.7 &
    1.50 &
    1.70 &
    \textbf{54.0} & \textbf{64.4} & \textbf{65.2} &
    3.85 & \textbf{3.91} & \textbf{3.94} &
    \textbf{79.0} & 89.2 & 10.3 \\
    \quad w/o inter-timings &
    24.5 &
    66.0 &
    3.20 &
    38.1 & 34.0 & 14.2 &
    3.54 & 3.49 & 2.67 &
    69.2 & 87.5 & 48.4 \\
    \bottomrule
  \end{tabular}
  \label{tab:main}
\end{table*}

\subsection{Base Audio Post-Training Data}

Base audio post-training is performed on a heterogeneous SFT mixture covering captioning, multi-turn dialogue, emotion recognition, and general voice tasks. Captioning data spans general-domain, multilingual, and music audio with both descriptive and QA-style supervision. Dialogue data includes speech-context conversations and interruptions. Emotion datasets provide discrete affect labels for acted speech. General tasks include multilingual speech recognition, multi-speaker processing, spotting, and TTS-style supervision. We keep this base mixture fixed across time-aware ablations to isolate the effect of temporal design choices.

\subsection{Time-aware Tasks}

We consider three time-centric generation tasks.

\textbf{Temporal grounding} asks the model to localize a text-described event by predicting a time interval in the audio; this is closely related to audio moment retrieval \cite{time_audio, amr}.

\textbf{Fragment description} conditions on a provided interval and produces a natural-language description of what happens in that window; this is analogous to audio captioning conditioned on a pre-segmented clip \cite{clotho, audiocaps}, and to speech summary settings when applied to long audio \cite{time_audio}.

\textbf{Summarization with timings} produces a multi-part summary anchored in time; unlike fragment description, the model must also \emph{decide the fragmentation} (i.e., choose segment boundaries) rather than receiving them as input.

\subsection{Audio and timestamped transcripts}

We start from six English shards of YODAS2 (24k hours) \cite{yodas} and filter to English with a SpeechBrain VoxLingua107 ECAPA-TDNN language-ID model, keeping only samples with $p(\text{English}) > 0.7$ (16k hours after filtering) \cite{speechbrain, voxlingua}. We then obtain word-level timestamps and time-aligned transcripts using WhisperX; the same alignment is also used to estimate the silence ratio \cite{whisperx}. Finally, we filter recordings by silence-ratio thresholds (keeping 14k hours), bucket them by duration ($\leq$20 min, 20--40 min, $\geq$40 min), and balance bucket sizes.

\subsection{Data generation} \label{subsec:data_gen}

Figure~\ref{fig:pipeline} summarizes the LLM components of our synthetic generation, illustrated using the temporal grounding task as an example. Long recordings are handled by generating QA from $\sim$10-minute transcript slices to reduce a strong \emph{front-loading bias} in question generation (without slicing, questions concentrate on the beginning of the recording). We generate synthetic data with a text-only GPT-OSS-120B \cite{gptoss} over timestamped transcripts, as audio-conditioned generation (audio+text input) performed worse in our preliminary experiments. A separate verifier consumes the full timestamped transcript and filters inconsistent generations. The train/eval split is done at the audio level. For the evaluation set, we additionally increase the sampling temperature for a fixed question, generate five answers, and filter out questions using a threshold on the median overlap between predicted time intervals, in order to remove incorrect examples while retaining challenging ones. For the other tasks, the aggregation stage is LLM-based. In particular, for summarization with timings, fragments are not sampled: the full timestamped transcript is provided to the generator in a single pass.

\subsection{Evaluation} \label{subsec:eval_focus}

We focus on \emph{temporal grounding} because it is the most verifiable: predictions can be scored by interval overlap without subjective judgments. We report (i) \emph{mIoU} between the predicted and reference intervals and (ii) \emph{MAE} (median absolute error in seconds). For interval outputs, MAE is computed at the midpoint $(\text{start} + \text{end})/2$. For fragment descriptions and timed summaries, standard reference-based automatic metrics are known to correlate imperfectly with human judgments \cite{study_sum}. Therefore, we evaluate these two tasks using an LLM-as-a-judge protocol that compares the model's prediction to a reference answer, with access to the full timestamped transcript.

For \emph{fragment descriptions}, the judge also receives the user question and evaluates (i) \emph{factual accuracy} (coverage of reference facts) and (ii) \emph{extra or contradictory information} (hallucinated or conflicting statements). An \emph{overall grade} is calculated from these criteria and reported as the final score on a 1-5 scale.

For \emph{summarization with timings}, we evaluate four dimensions: (1) \emph{Tm (timing structure)} - correctness of chronological order, quality and balance of segmentation, and semantic alignment of time blocks; (2) \emph{Acc (factual accuracy)} - computed as a weighted coverage ratio over reference facts (key and secondary assigned different importance weights); (3) \emph{Err (errors and contradictions)} - a weighted error ratio over candidate facts proportionally to their importance; and (4) \emph{Style (structure and style)} - logical organization, coherence, conciseness, and formatting quality. Additionally, for timing evaluation we report an automatic metric measuring the proportion of \emph{round segments}. A segment is considered round if its start and end timestamps are a whole number of minutes apart (e.g., \texttt{12:23-14:23}); a high proportion of such segments may indicate coarse segmentation.

\section{Experiments}

\subsection{Main comparison across tasks and datasets} \label{subsec:main_comp}

Table~\ref{tab:main} summarizes performance across four benchmarks:

\begin{enumerate}[label=(\roman*)]

\item \textsc{AudioGrounding}~\cite{audiogrounding}. 
Short-clip dataset (980 samples, 7--10\,s). 
Questions refer to acoustic events, and the model predicts a temporal interval. 
We report mIoU.

\item \textsc{AMI Meeting Corpus}~\cite{ami_meeting}. 
Meeting recordings (15--50\,min) with word-level timestamps. 
We automatically generate 150 phrase-level questions targeting 3--5\,s utterances (``when was this phrase spoken?''). 
We report interval MAE.

\item \textsc{DCASE Audio QA} (time-aware subset)~\cite{dcase_qa}.
Multiple-choice dataset; we select questions of the form 
\texttt{What is the start time\ldots}, resulting in 144 samples (6-10\,s) with time answers such as \texttt{5.2s}. 
We report pointwise MAE.

\item Our benchmarks covering three tasks. 
For temporal grounding we report mIoU; for fragment description we use LLM-as-a-judge overall score metrics (see Section~\ref{subsec:eval_focus}). For timed summarization, we report three metrics: \emph{Tm} - judge-based timing structure score; \emph{AES} - aggregate content score computed as $(\text{Acc} + 1 - \text{Err} + \text{Style}) / 3$, where \emph{Acc}, \emph{Err}, and \emph{Style} are the corresponding dimensions of the LLM judge described above; \emph{Rd} - share of round segments, i.e., segments whose duration is a multiple of 60 seconds. Results are bucketed by duration (0--1\,min, 2--5\,min, 20--40\,min); summarization is evaluated only on the longest bucket.

\end{enumerate}

We include open multimodal baselines where available. In our internal evaluation, TimeAudio \cite{time_audio} degrades on audio longer than two minutes, frequently producing \texttt{UNK} and non-parseable timestamps. We therefore report its results only when the benchmark duration matches its original setup. Table~\ref{tab:main} highlights that strong short-clip performance does not translate to long-form temporal grounding (e.g., Qwen3-Omni drops to 3.6 mIoU on 20--40m TGr and 290.5\,s interval MAE on AMI). In contrast, our model maintains stable grounding accuracy as duration increases (53.8 mIoU on 20--40\,min) and further improves when inserting temporal anchors every 7 seconds (65.2 mIoU). The gap to our own ablation confirms that \emph{inter-timings are critical}: removing periodic temporal anchors reduces long-form mIoU from 53.8 to 14.2 and degrades timed descriptions and summaries. In the following, we systematically study the optimal frequency and textual encoding of inter-timing anchors.

\subsection{Length extrapolation}

\begin{figure}[t]
  \centering
  \includegraphics[width=\linewidth]{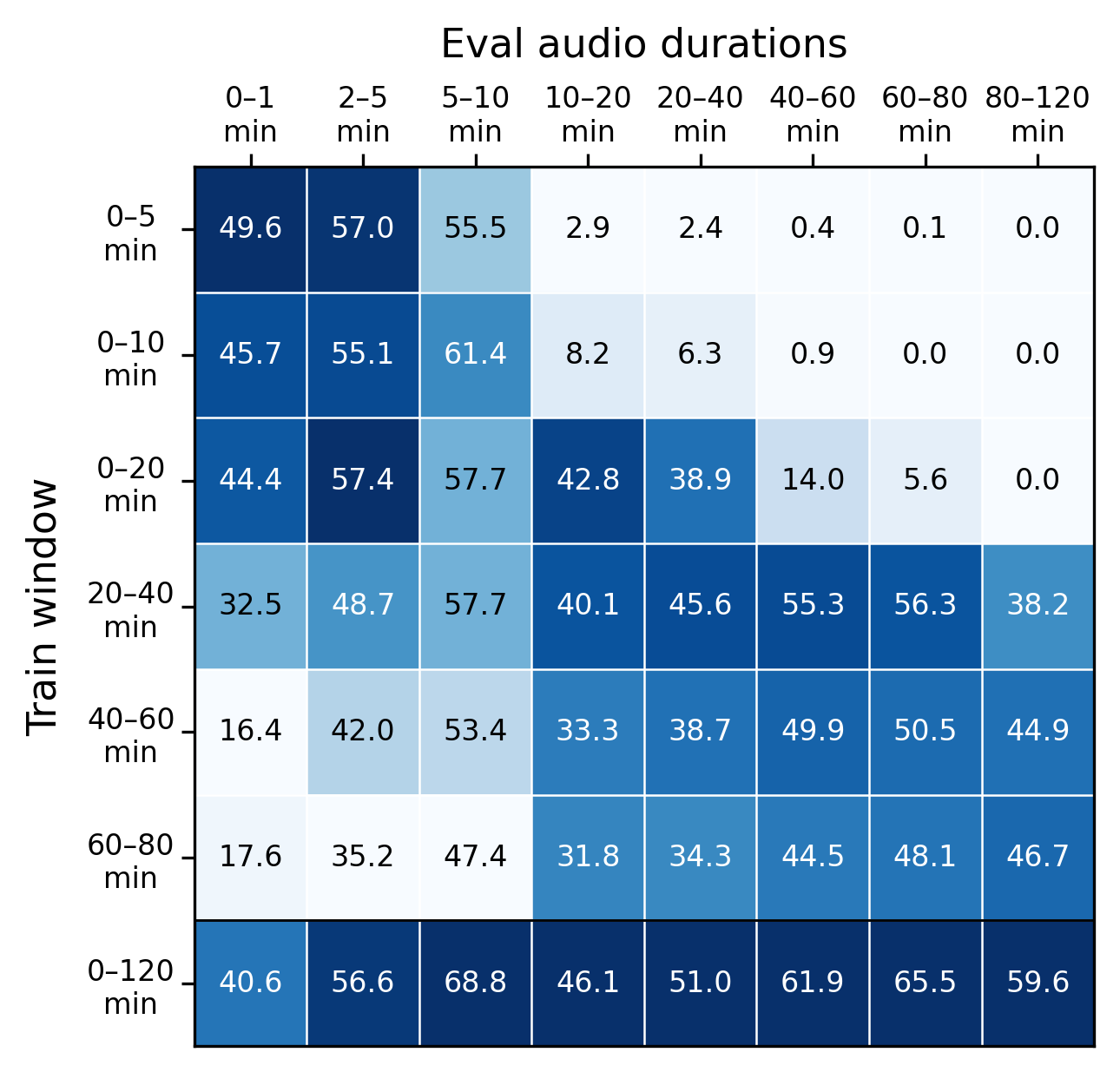}
  \caption{
    \textbf{Length extrapolation up to 120\,min for temporal grounding in mIoU ($\uparrow$)}. Rows are training duration windows, columns are eval audio lengths. Shading is column-normalized, indicating relative quality within each eval duration.
  }
  \label{fig:extrap}
\end{figure}

We evaluate how models trained on specific duration regimes generalize across lengths up to 120 minutes (Figure~\ref{fig:extrap}). Checkpoints are trained on different duration buckets (rows) and evaluated on a shared duration grid (columns). We observe asymmetric generalization: training only on short audio fails on long recordings, while training only on long audio hurts short-audio performance. To ensure clean separation, we discard early-recording questions (e.g., for 10–20\,min evaluation, answers must occur after minute 10). The model trained on all durations outperforms other regimes across nearly all audio lengths.

\subsection{Special timing tokens and scaling the timing ratio} \label{subsec:extratokens}

We test whether introducing dedicated timestamp tokens improves grounding. We keep the surface format \texttt{hh:mm:ss} but replace its characters with \emph{special timing tokens} initialized from the embeddings of corresponding digits and the colon token. Table~\ref{tab:extratokens} shows that these extra tokens require a much larger timing-token share in SFT to match the plain-text baseline; only at very high timing ratios do they reach comparable mIoU.

\begin{table}[t]
  \caption{
      \textbf{Special timing tokens vs. timings data ratio.}
      We report mIoU for models trained with plain-text (regular) and with special (extra) timing tokens, evaluated at different ratios of temporal grounding (TG) data within the audio SFT mixture.
  }
  \centering
  \small
  \setlength{\tabcolsep}{4pt}
  \begin{tabular}{lcccccc}
    \toprule
    \textbf{TG ratio} & 2.4\% & 4.8\% & 9.1\% & 16.7\% & 33.4\% & 50.0\% \\
    \midrule
    regular      & 41.4 & 50.9 & 52.5 & 57.5 & --   & --   \\
    extra tokens & --   & 13.1 & 37.6 & 50.0 & 55.2 & 56.8 \\
    \bottomrule
  \end{tabular}
  \label{tab:extratokens}
\end{table}

\subsection{Inter-timing frequency} \label{subsec:interfreq}

Inter-timings act as periodic anchors inserted into the input stream. Table~\ref{tab:freq} shows that more frequent anchors improve grounding, at the cost of additional tokens.
Even with 60\,s anchors, the model achieves a median error of 3\,s (vs.\ 1.5\,s for 7\,s anchors), effectively interpolating second-level precision between minute-spaced anchors.

\begin{table}[t]
  \caption{
    \textbf{Inter-timing frequency trade-off.}
    mIoU and MAE (in seconds, on interval midpoints) are reported for different anchor intervals.
    The added ratio shows the share of timing tokens relative to audio tokens at a 160\,ms frame rate.
  }
  \centering
  \small
  \setlength{\tabcolsep}{6pt}
  \begin{tabular}{lcccccc}
    \toprule
    \textbf{Freq} & 7s & 15s & 30s & 60s & 120s & 240s \\
    \midrule
    mIoU ($\uparrow$)  & 63.0 & 59.9 & 55.5 & 50.9 & 41.3 & 31.0 \\
    MAE ($\downarrow$) & 1.5 & 2.0 & 2.5 & 3.0 & 5.0 & 8.5 \\
    ad. ratio & 16.0\% & 7.5\% & 3.7\% & 1.9\% & 0.9\% & 0.5\% \\
    \bottomrule
  \end{tabular}
  \label{tab:freq}
\end{table}

\subsection{Inter-timing format}

With one anchor per minute, the canonical \texttt{hh:mm:ss} format can be token-heavy (e.g., \texttt{00:01:00}). We compare alternative encodings that shorten the marker without changing its semantics. Table~\ref{tab:format} shows that pure seconds degrade performance substantially, while minute-index formats remain close to \texttt{hh:mm:ss} with much lower token overhead.

\begin{table}[!t]
  \caption{
    \textbf{Inter-timing marker format (one marker per 60s).}
    We report temporal grounding performance for different textual encodings of the inter-timing marker.
  }
  \centering
  \small
  \setlength{\tabcolsep}{6pt}
  \begin{tabular}{lccccc}
    \toprule
    \textbf{Format} &
    \texttt{hh:mm:ss} &
    \texttt{m:0} &
    \texttt{m:} &
    \texttt{m} &
    \texttt{sec} \\
    \midrule
    mIoU & 50.9 & 47.1 & 43.8 & 44.5 & 20.9 \\
    \bottomrule
  \end{tabular}
  \label{tab:format}
\end{table}

\section{Conclusion}

We presented a time-aware Audio LLM that supports up to 120 minutes of input and produces answers and summaries explicitly anchored in time. Our results show that temporal grounding in long recordings remains a major bottleneck for existing multimodal models, which often degrade sharply beyond a few minutes of audio. We demonstrate that periodic temporal anchors (inter-timings) are essential for stable long-form grounding, while even sparse anchors (e.g., once per minute) are sufficient for reliable grounding, and that training on a mixture of audio durations is necessary for length generalization. Alongside the model, we release a 10k+ hours dataset spanning seconds-to-hours recordings for temporal QA and timed summarization, aiming to facilitate further research on time-aware audio understanding.

\clearpage

\section{Generative AI Use Disclosure}

Generative AI tools were used for limited language editing (grammar, clarity, and stylistic refinement) during manuscript preparation. All outputs were carefully reviewed, edited, and validated by the authors, who take full responsibility for the final content. As described in Section~\ref{subsec:data_gen} and Section~\ref{subsec:eval_focus}, large language models were used within the experimental framework: (i) to generate synthetic supervision from timestamped transcripts, and (ii) as LLM-as-a-judge evaluators for fragment description and timed summarization. Generative AI tools did not formulate scientific claims, interpret results, or determine research conclusions.

\bibliographystyle{IEEEtran}
\bibliography{mybib}

% Generated by IEEEtran.bst, version: 1.13 (2008/09/30)
\begin{thebibliography}{10}
\providecommand{\url}[1]{#1}
\csname url@samestyle\endcsname
\providecommand{\newblock}{\relax}
\providecommand{\bibinfo}[2]{#2}
\providecommand{\BIBentrySTDinterwordspacing}{\spaceskip=0pt\relax}
\providecommand{\BIBentryALTinterwordstretchfactor}{4}
\providecommand{\BIBentryALTinterwordspacing}{\spaceskip=\fontdimen2\font plus
\BIBentryALTinterwordstretchfactor\fontdimen3\font minus \fontdimen4\font\relax}
\providecommand{\BIBforeignlanguage}[2]{{%
\expandafter\ifx\csname l@#1\endcsname\relax
\typeout{** WARNING: IEEEtran.bst: No hyphenation pattern has been}%
\typeout{** loaded for the language `#1'. Using the pattern for}%
\typeout{** the default language instead.}%
\else
\language=\csname l@#1\endcsname
\fi
#2}}
\providecommand{\BIBdecl}{\relax}
\BIBdecl

\bibitem{gpt4o}
\BIBentryALTinterwordspacing
OpenAI, ``Gpt-4o system card,'' 2024. [Online]. Available: \url{https://arxiv.org/abs/2410.21276}
\BIBentrySTDinterwordspacing

\bibitem{gemini2_5}
\BIBentryALTinterwordspacing
{Gemini Team, Google}, ``Gemini 2.5: Pushing the frontier with advanced reasoning, multimodality, long context, and next generation agentic capabilities,'' 2025. [Online]. Available: \url{https://arxiv.org/abs/2507.06261}
\BIBentrySTDinterwordspacing

\bibitem{voxtral}
\BIBentryALTinterwordspacing
A.~H. Liu \emph{et~al.}, ``Voxtral,'' 2025. [Online]. Available: \url{https://arxiv.org/abs/2507.13264}
\BIBentrySTDinterwordspacing

\bibitem{qwen3omni}
\BIBentryALTinterwordspacing
J.~Xu \emph{et~al.}, ``Qwen3-omni technical report,'' 2025. [Online]. Available: \url{https://arxiv.org/abs/2509.17765}
\BIBentrySTDinterwordspacing

\bibitem{qwen2_audio}
\BIBentryALTinterwordspacing
Y.~Chu, J.~Xu, Q.~Yang, H.~Wei, X.~Wei, Z.~Guo, Y.~Leng, Y.~Lv, J.~He, J.~Lin, C.~Zhou, and J.~Zhou, ``Qwen2-audio technical report,'' 2024. [Online]. Available: \url{https://arxiv.org/abs/2407.10759}
\BIBentrySTDinterwordspacing

\bibitem{salmonn}
\BIBentryALTinterwordspacing
C.~Tang, W.~Yu, G.~Sun, X.~Chen, T.~Tan, W.~Li, L.~Lu, Z.~Ma, and C.~Zhang, ``{SALMONN}: Towards generic hearing abilities for large language models,'' 2024. [Online]. Available: \url{https://arxiv.org/abs/2310.13289}
\BIBentrySTDinterwordspacing

\bibitem{vibe_voice_asr}
\BIBentryALTinterwordspacing
Z.~Peng \emph{et~al.}, ``{VIBEVOICE-ASR} technical report,'' 2026. [Online]. Available: \url{https://arxiv.org/abs/2601.18184}
\BIBentrySTDinterwordspacing

\bibitem{moss_transcribe}
\BIBentryALTinterwordspacing
MOSI.AI, ``Moss transcribe diarize technical report,'' 2026. [Online]. Available: \url{https://arxiv.org/abs/2601.01554}
\BIBentrySTDinterwordspacing

\bibitem{whisperx}
M.~Bain, J.~Huh, T.~Han, and A.~Zisserman, ``{WhisperX: Time-Accurate Speech Transcription of Long-Form Audio},'' in \emph{{Interspeech 2023}}, 2023, pp. 4489--4493.

\bibitem{audiogrounding}
X.~Xu, H.~Dinkel, M.~Wu, and K.~Yu, ``Text-to-audio grounding: Building correspondence between captions and sound events,'' in \emph{ICASSP 2021 - 2021 IEEE International Conference on Acoustics, Speech and Signal Processing (ICASSP)}, 2021, pp. 606--610.

\bibitem{dcase_qa}
\BIBentryALTinterwordspacing
C.-H.~H. Yang, S.~Ghosh, Q.~Wang, J.~Kim, H.~Hong, S.~Kumar, G.~Zhong, Z.~Kong, S.~Sakshi, V.~Lokegaonkar, O.~Nieto, R.~Duraiswami, D.~Manocha, G.~Kim, J.~Du, R.~Valle, and B.~Catanzaro, ``Multi-domain audio question answering toward acoustic content reasoning in the dcase 2025 challenge,'' 2025. [Online]. Available: \url{https://arxiv.org/abs/2505.07365}
\BIBentrySTDinterwordspacing

\bibitem{time_audio}
\BIBentryALTinterwordspacing
H.~Wang, Y.~Li, S.~Ma, H.~Liu, and X.~Wang, ``Listening between the frames: Bridging temporal gaps in large audio-language models,'' 2025. [Online]. Available: \url{https://arxiv.org/abs/2511.11039}
\BIBentrySTDinterwordspacing

\bibitem{amr}
H.~Munakata, T.~Nishimura, S.~Nakada, and T.~Komatsu, ``Language-based audio moment retrieval,'' in \emph{ICASSP 2025 - 2025 IEEE International Conference on Acoustics, Speech and Signal Processing (ICASSP)}, 2025, pp. 1--5.

\bibitem{castella}
\BIBentryALTinterwordspacing
H.~Munakata, T.~Imamura, T.~Nishimura, and T.~Komatsu, ``{CASTELLA}: Long audio dataset with captions and temporal boundaries,'' 2026. [Online]. Available: \url{https://arxiv.org/abs/2511.15131}
\BIBentrySTDinterwordspacing

\bibitem{video_moment_localization}
\BIBentryALTinterwordspacing
L.~A. Hendricks, O.~Wang, E.~Shechtman, J.~Sivic, T.~Darrell, and B.~Russell, ``Localizing moments in video with temporal language,'' in \emph{Proceedings of the 2018 Conference on Empirical Methods in Natural Language Processing}, E.~Riloff, D.~Chiang, J.~Hockenmaier, and J.~Tsujii, Eds.\hskip 1em plus 0.5em minus 0.4em\relax Brussels, Belgium: Association for Computational Linguistics, Oct.-Nov. 2018, pp. 1380--1390. [Online]. Available: \url{https://aclanthology.org/D18-1168/}
\BIBentrySTDinterwordspacing

\bibitem{timemarker}
\BIBentryALTinterwordspacing
S.~Chen, X.~Lan, Y.~Yuan, Z.~Jie, and L.~Ma, ``Timemarker: A versatile video-llm for long and short video understanding with superior temporal localization ability,'' 2024. [Online]. Available: \url{https://arxiv.org/abs/2411.18211}
\BIBentrySTDinterwordspacing

\bibitem{longcat}
\BIBentryALTinterwordspacing
M.~L. Team, Bayan, B.~Li, and B.~L. et~al., ``Longcat-flash technical report,'' 2025. [Online]. Available: \url{https://arxiv.org/abs/2509.01322}
\BIBentrySTDinterwordspacing

\bibitem{geval}
\BIBentryALTinterwordspacing
Y.~Liu, D.~Iter, Y.~Xu, S.~Wang, R.~Xu, and C.~Zhu, ``{G}-eval: {NLG} evaluation using gpt-4 with better human alignment,'' in \emph{Proceedings of the 2023 Conference on Empirical Methods in Natural Language Processing}, H.~Bouamor, J.~Pino, and K.~Bali, Eds.\hskip 1em plus 0.5em minus 0.4em\relax Singapore: Association for Computational Linguistics, Dec. 2023, pp. 2511--2522. [Online]. Available: \url{https://aclanthology.org/2023.emnlp-main.153/}
\BIBentrySTDinterwordspacing

\bibitem{10b}
{ai-sage}, ``{GigaChat3-10B-A1.8B},'' \url{https://huggingface.co/ai-sage/GigaChat3-10B-A1.8B}, 2025, hugging Face model card; accessed 2026-02-25.

\bibitem{flash}
\BIBentryALTinterwordspacing
T.~Dao, ``{FlashAttention-2}: Faster attention with better parallelism and work partitioning,'' in \emph{The Twelfth International Conference on Learning Representations}, 2024. [Online]. Available: \url{https://openreview.net/forum?id=mZn2Xyh9Ec}
\BIBentrySTDinterwordspacing

\bibitem{hubert}
W.-N. Hsu, B.~Bolte, Y.-H.~H. Tsai, K.~Lakhotia, R.~Salakhutdinov, and A.~Mohamed, ``Hubert: Self-supervised speech representation learning by masked prediction of hidden units,'' \emph{IEEE/ACM Transactions on Audio, Speech, and Language Processing}, vol.~29, pp. 3451--3460, 2021.

\bibitem{gigaam}
A.~Kutsakov, A.~Maximenko, G.~Gospodinov, P.~Bogomolov, and F.~Minkin, ``{GigaAM: Efficient Self-Supervised Learner for Speech Recognition},'' in \emph{{Interspeech 2025}}, 2025, pp. 1213--1217.

\bibitem{moe}
\BIBentryALTinterwordspacing
N.~Shazeer, A.~Mirhoseini, K.~Maziarz, A.~Davis, Q.~Le, G.~Hinton, and J.~Dean, ``Outrageously large neural networks: The sparsely-gated mixture-of-experts layer,'' 2017. [Online]. Available: \url{https://arxiv.org/abs/1701.06538}
\BIBentrySTDinterwordspacing

\bibitem{clotho}
K.~Drossos, S.~Lipping, and T.~Virtanen, ``Clotho: an audio captioning dataset,'' in \emph{ICASSP 2020 - 2020 IEEE International Conference on Acoustics, Speech and Signal Processing (ICASSP)}, 2020, pp. 736--740.

\bibitem{audiocaps}
\BIBentryALTinterwordspacing
C.~D. Kim, B.~Kim, H.~Lee, and G.~Kim, ``{A}udio{C}aps: Generating captions for audios in the wild,'' in \emph{Proceedings of the 2019 Conference of the North {A}merican Chapter of the Association for Computational Linguistics: Human Language Technologies, Volume 1 (Long and Short Papers)}, J.~Burstein, C.~Doran, and T.~Solorio, Eds.\hskip 1em plus 0.5em minus 0.4em\relax Minneapolis, Minnesota: Association for Computational Linguistics, Jun. 2019, pp. 119--132. [Online]. Available: \url{https://aclanthology.org/N19-1011/}
\BIBentrySTDinterwordspacing

\bibitem{yodas}
X.~Li, S.~Takamichi, T.~Saeki, W.~Chen, S.~Shiota, and S.~Watanabe, ``\BIBforeignlanguage{English}{Yodas: Youtube-oriented dataset for audio and speech},'' in \emph{\BIBforeignlanguage{English}{2023 IEEE Automatic Speech Recognition and Understanding Workshop, ASRU 2023}}, ser. 2023 IEEE Automatic Speech Recognition and Understanding Workshop, ASRU 2023.\hskip 1em plus 0.5em minus 0.4em\relax Institute of Electrical and Electronics Engineers Inc., 2023, publisher Copyright: {\textcopyright} 2023 IEEE.; 2023 IEEE Automatic Speech Recognition and Understanding Workshop, ASRU 2023 ; Conference date: 16-12-2023 Through 20-12-2023.

\bibitem{speechbrain}
\BIBentryALTinterwordspacing
T.~Parcollet, M.~Ravanelli, P.~Plantinga, A.~Rouhe, S.~Cornell, L.~Lugosch, C.~Subakan, N.~Dawalatabad, A.~Heba, J.~Zhong, J.-C. Chou, S.-L. Yeh, S.-W. Fu, C.-F. Liao, E.~Rastorgueva, F.~Grondin, W.~Aris, H.~Na, Y.~Gao, R.~de~Mori, and Y.~Bengio, ``{SpeechBrain: A General-Purpose Speech Toolkit},'' Mar. 2022, preprint. [Online]. Available: \url{https://hal.science/hal-03601303}
\BIBentrySTDinterwordspacing

\bibitem{voxlingua}
\BIBentryALTinterwordspacing
J.~Valk and T.~Alum{\"a}e, ``Voxlingua107: A dataset for spoken language recognition,'' \emph{2021 IEEE Spoken Language Technology Workshop (SLT)}, pp. 652--658, 2020. [Online]. Available: \url{https://api.semanticscholar.org/CorpusID:227209238}
\BIBentrySTDinterwordspacing

\bibitem{gptoss}
\BIBentryALTinterwordspacing
OpenAI, S.~Agarwal, L.~Ahmad, J.~Ai, S.~Altman, A.~Applebaum \emph{et~al.}, ``gpt-oss-120b \& gpt-oss-20b model card,'' 2025. [Online]. Available: \url{https://arxiv.org/abs/2508.10925}
\BIBentrySTDinterwordspacing

\bibitem{study_sum}
\BIBentryALTinterwordspacing
H.~Nguyen, H.~Chen, L.~Pobbathi, and J.~Ding, ``A comparative study of quality evaluation methods for text summarization,'' \emph{ArXiv}, vol. abs/2407.00747, 2024. [Online]. Available: \url{https://api.semanticscholar.org/CorpusID:270869494}
\BIBentrySTDinterwordspacing

\bibitem{summeval}
\BIBentryALTinterwordspacing
A.~R. Fabbri, W.~Kryscinski, B.~McCann, R.~Socher, and D.~R. Radev, ``Summeval: Re-evaluating summarization evaluation,'' \emph{Transactions of the Association for Computational Linguistics}, vol.~9, pp. 391--409, 2020. [Online]. Available: \url{https://api.semanticscholar.org/CorpusID:220768873}
\BIBentrySTDinterwordspacing

\bibitem{ami_meeting}
J.~Carletta, ``\BIBforeignlanguage{English}{Unleashing the killer corpus: experiences in creating the multi-everything ami meeting corpus},'' \emph{\BIBforeignlanguage{English}{Language Resources and Evaluation}}, vol.~41, no.~2, pp. 181--190, 2007.

\end{thebibliography}

\end{document}